# Ultra-High Performance of Nano-Engineered Graphene-Based Natural Jute Fiber Composites

Forkan Sarker,[1,2] Prasad Potluri,[1,2]* Shaila Afroj,[3,4] Vivek Koncherry,[1,2] Kostya S Novoselov,[3,4] and Nazmul Karim[4]*

[1]School of Materials, The University of Manchester, Oxford Road, Manchester, M13 9PL, UK

[2]Northwest Composites Centre, The University of Manchester, Oxford Road, Manchester, M13 9PL, UK

[3]School of Physics & Astronomy, The University of Manchester, Oxford Road, Manchester, M13 9PL, UK

[4]National Graphene Institute (NGI), The University of Manchester, Booth Street East, Manchester, M13 9PL, UK

*Corresponding E-mail: mdnazmul.karim@manchester.ac.uk and prasad.potluri@manchester.ac.uk;

**Abstract**

*Natural fibers composites are considered as sustainable alternative to synthetic composites due to their environmental and economic benefits. However, they suffer from poor mechanical and interfacial properties due to a random fiber orientation and weak fiber-matrix interface. Here we report nano-engineered graphene-based natural jute fiber preforms with a new fiber architecture (NFA) which significantly improves their properties and performances. Our graphene-based NFA of jute fiber perform enhances Young modulus of jute-epoxy composites by ~324% and tensile strength by ~110% more than untreated jute fiber composites, by arranging fibers in parallel direction through individualisation and nano surface engineering with graphene derivatives. This could potentially lead to manufacturing of high performance natural alternatives to synthetic composites in various stiffness driven high performance applications.*

Natural fiber reinforced composites (FRC) have been a focus of much attention over the recent years due to their potential to replace environmentally unfriendly synthetic FRC.[1] Moreover, natural fibers comes from the renewable resource and can easily be recycled or burned with less residue and $CO_2$ emission to the atmosphere.[1, 2] Such lightweight and environmentally sustainable FRC could ideally be used as a replacement of glass, carbon or other synthetic FRC,[3] in numerous applications such as automotive, construction and household.[4] Jute, flax, hemp and sisal are the main dominating natural bast fibers that are used as reinforcing materials for FRC. Among them, jute fibers have been a popular choice as reinforcing materials for composites due to lower production cost, lower density, long individual fiber length and better mechanical properties than other natural alternatives.[5] However, jute FRC still suffer from poor mechanical properties, when compared with synthetic fibers (such as glass).[6]

The mechanical properties of FRC are mainly dominated by: a) The properties of reinforcing materials which are considered to be the main load bearing constituents and b) The interface between the fiber and matrix which transfer the load from the matrix to the reinforcing materials through shear stress.[7, 8] Alkali treatment is a popular technique to enhance the mechanical properties of natural FRC by improving the load bearing capacity of reinforcing materials and creating strong fiber-matrix interface.[9, 10] It removes impurities such as wax, hemi-cellulose and lignin from the fiber surface, and separates elementary fibers from technical fibers in order to improve the fiber packing in composites. As suggested in the previous studies,[6, 11-13] the treatment with a lower alkali concentration (~0.5 wt.-%) for a prolonged time is an effective technique to enhance the mechanical properties of jute fiber. However, there are still flaws[5] (micro-voids) present in the fiber even after the alkali treatment that inhibits fibrils to carry more loads and produce weak fiber/matrix interface. There have been many efforts to enhance the performance of composites by removing these flaws either through nano materials grafting[7, 14-18] or surface modifications such as silane treatment,[19] acetylations,[20] etherification,[21] peroxide[22] and plasma treatment.[23] However, such treatments are expensive and time consuming process with very limited improvement in composites performances.

Graphene-based materials have shown huge potential for composite applications due to their excellent mechanical properties. Graphene Oxide (GO) is a graphene derivative, which is formed by attaching various oxygen functional groups (e.g. hydroxyl group, epoxy and carbonyl group) to their basal panel and edges of a graphene sheet.[15, 24, 25] Therefore, GO could add functionality to fibers and enhance the strength and toughness at fiber/matrix interface.[26] Several studies reported the use of GO to improve the performance of synthetic composites such as grafting of GO onto glass fibers[7] and carbon fabric,[18] and blending of GO into epoxy resin for carbon/epoxy composites.[27] Moreover, Graphene Nanoplatelets (GNP) or graphene flakes (G) have been investigated for composite applications,[28-32] as such materials could be produced in large quantity.[33] The study suggested that GNP could prevent the delamination of the fibers[31] and can delay the crack propagation at the interphase by redistributing the stress around fibers, where the cracks started to from.[18] However, very limited study has been carried out on natural fiber-based composites for structural applications. In our previous study, we report that the coating of graphene materials (GO and G flakes) onto jute fibers enhanced interfacial shear strength and tensile strength of individual fibers by ~236% and ~96%, respectively.[6] However, the main challenge is how we could translate such excellent properties achieved on individual fibers to a jute fiber reinforced composite for real world applications.

Here we address this challenge by reporting a novel strategy to manufacture next generation natural fiber reinforced composites by combining physical and chemical modification of jute fibers preforms. A simple hand combing was used to individualise jute fibers with subsequent alkali treatment to remove non-cellulosic impurities from jute fiber surface. Then jute fiber was modified by GO and G flakes with subsequent hot pressing to produce preforms with a new fiber architecture (NFA), before jute/epoxy composites were made by a vacuum resin infusion process. The improvement in longitudinal and transverse mechanical properties of composites with surface treatment and NFA of jute fiber preforms was tested using a tensile tester. The fracture surface of tested specimen was analysed using a Scanning Electron Microscope (SEM). Finally, obtained tensile and specific properties of as prepared jute fibers composites were compared with that of glass and flax fibers and also with results reported in literatures for surface modified natural fibers.

**Results and Discussion**

**Improved Fiber Volume Fraction with Physical and Chemical Treatments**

The fiber volume fraction ($V_f$) of FRC have significant effect on mechanical properties (such as strength, stiffness and toughness) of the composites materials.[34] The strength and stiffness of a composite laminate increases proportionally with the increase of $V_f$ upto ~80%, at which the amount of resin is sufficient to hold the fibers properly.[35] However, jute fiber composites suffer from relatively lower $V_f$ (~23%) during the vacuum infusion process like other natural fiber composites, may be due to the presence of impurities and inter-fibrillar arrangement in the fiber. Moreover, the presence of waxes, lignin, and hemicellulose in jute fibers provides very smooth fiber surface (*Figure S2a, Supporting Information*) and do not allow fibrils to come out and pack inside the composites, Figure 1a. Therefore, lower $V_f$ value is obtained, which usually results in jute fiber reinforced composites with poor failure mode and ultimate strength.[36]

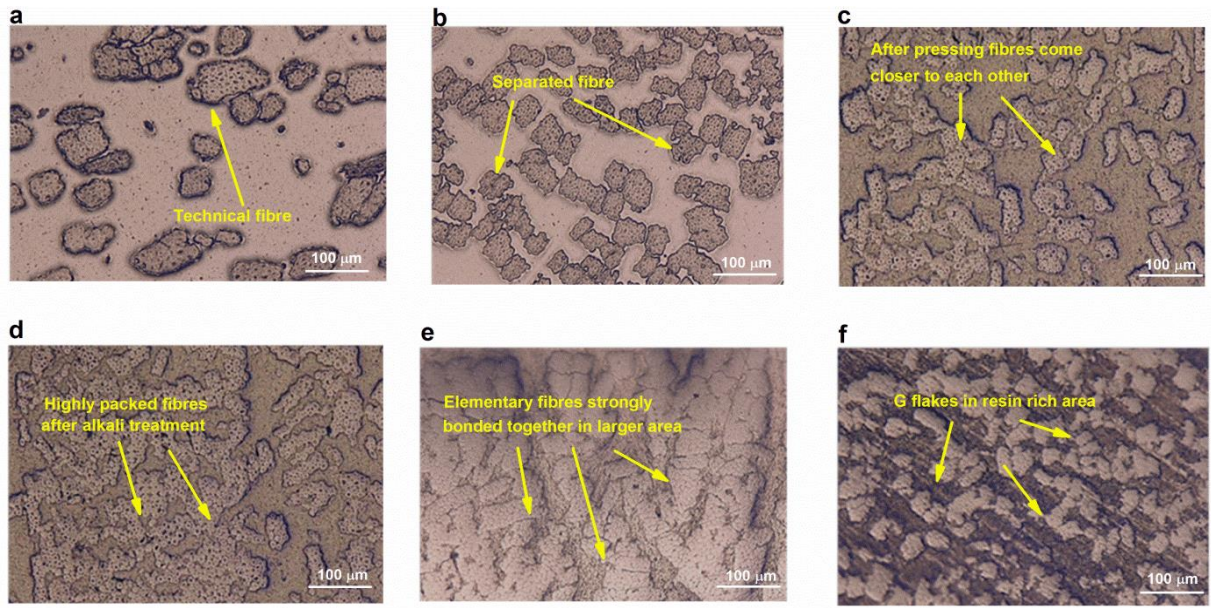

*Figure 1.* *Optical microscopic cross-sectional image of jute fiber epoxy composites: (a) untreated fiber composites, UT (X500); (b) fibrillated jute fiber composites, NFA (X500); (c) NFA composites with pressing, NFAHP (X500); (d) alkali treated composites HA0.5 (X500); (e) GO coated composites GO 0.75 (X500); and (f) G flake coated composites G10 (X500).*

In order to improve the mechanical properties of jute FRC, here we develop jute fibre preforms with novel surface treatments and a new fiber architecture (NFA), that would increase $V_f$ significantly by exposing jute fibers to physical (combing and hot pressing) and chemical treatments (alkali and graphene-based materials). We perceive the increment of $V_f$ after each treatment by observing the cross-section of jute/epoxy composites under the optical microscope, Figure 1a-f. The combing of jute fibers increases $V_f$ by ~10% to ~33% due to the increase in the degree of fiber separation in the preform[37] and fiber packing in the composites as shown in Figure 1b. We then applied hot pressing on combed fiber in order to achieve a NFA that results in a significant increase of $V_f$ to ~47.5%, which is in agreement with previous studies.[38-40] They reported $V_f$ of ~60-70% with unidirectional natural fiber composites at a constant compaction pressure and ~35-40% for composites with randomly oriented fibers. However, all the processes mentioned in the literature are based on liquid moulding (i.e. hand lay-up, RTM). In contrast, we report for the first time that manufacturing of a compact dry jute fiber preform (*Figure S2b, Supporting Information*) with a new architecture in order to improve the fiber packing. The NFA will enable potential manufacturing of composites with complex structure and excellent drapability.

We then investigate the effects of chemical treatments on $V_f$ of jute/epoxy composites. The alkali (0.5 wt.-%) treatment of combed jute fibers and subsequent hot pressing further increases $V_f$ to ~54%. Alkali treatment removes hemicellulose present between the fibrils and improves the fiber packing within the jute/epoxy composites, Figure 1d. The coating with graphene-materials on combed and alkali treated fibers increases $V_f$ slightly to ~55% (with G Flakes) and ~56% (with GO) after compaction with a hot press. The coating with GO provides slightly better $V_f$ than that of G flakes coated fibers, may be due to strong bonding between oxygen containing functional groups of cellulosic fibers and that of GO, Figure 1e.[6, 24, 41] In contrast, G-flake coated jute fibers do not produce strong bonding due to the absence of oxygen functional group.[28] Moreover, Figure 1f shows agglomeration of G flakes around the fiber surface and matrix.

**Enhanced Mechanical Properties with NFA and Physical Treatments**

Jute fiber contain large amount of (20–50 wt.-%) of non-cellulosic materials such as hemicellulose and lignin. Such non-cellulosic materials are responsible for the lower crystallinity and hydrophilic nature of jute fibers. Moreover, UT jute fibers contain mostly technical fibers bundles (consists of individual elementary fibers, Figure 2a), which is found to be have ~41% less Young's modulus and ~39% less tensile strength than the individual elementary fiber (*Table S1, Supporting Information*). Furthermore, lower $V_f$ is obtained with UT jute fiber composites. Therefore jute fiber composites suffer from poor tensile properties when reinforced with epoxy matrix. We obtain lower Young modulus (~10 GPa) and tensile strength (~180 MPa) with untreated jute/epoxy composites, Figure 2d.

We therefore develop a new fiber architecture (NFA) for jute fiber preform by using a simple hand combing process, Figure 2b. This results in increment in individualised and homogenous elementary fibers with fewer defects and better mechanical properties.[42] The Young's modulus and tensile strength are increased by ~95% to ~20.5 GPa and ~12% to ~202 MPa, respectively for jute/epoxy composites with NFA (Figure 2(d, e) *and Table S2, Supporting Information*) than that of UT composites. Such a significant improvement in the mechanical properties of the composites could be associated with structural mechanics of reinforcing fibers and their increased load bearing capacity. Moreover, NFA fiber composites show higher fiber content ($V_f$) due to the combing process, whereas in UT fiber composites most of the fiber bundles agglomerated in the cross section, Figure 1a.

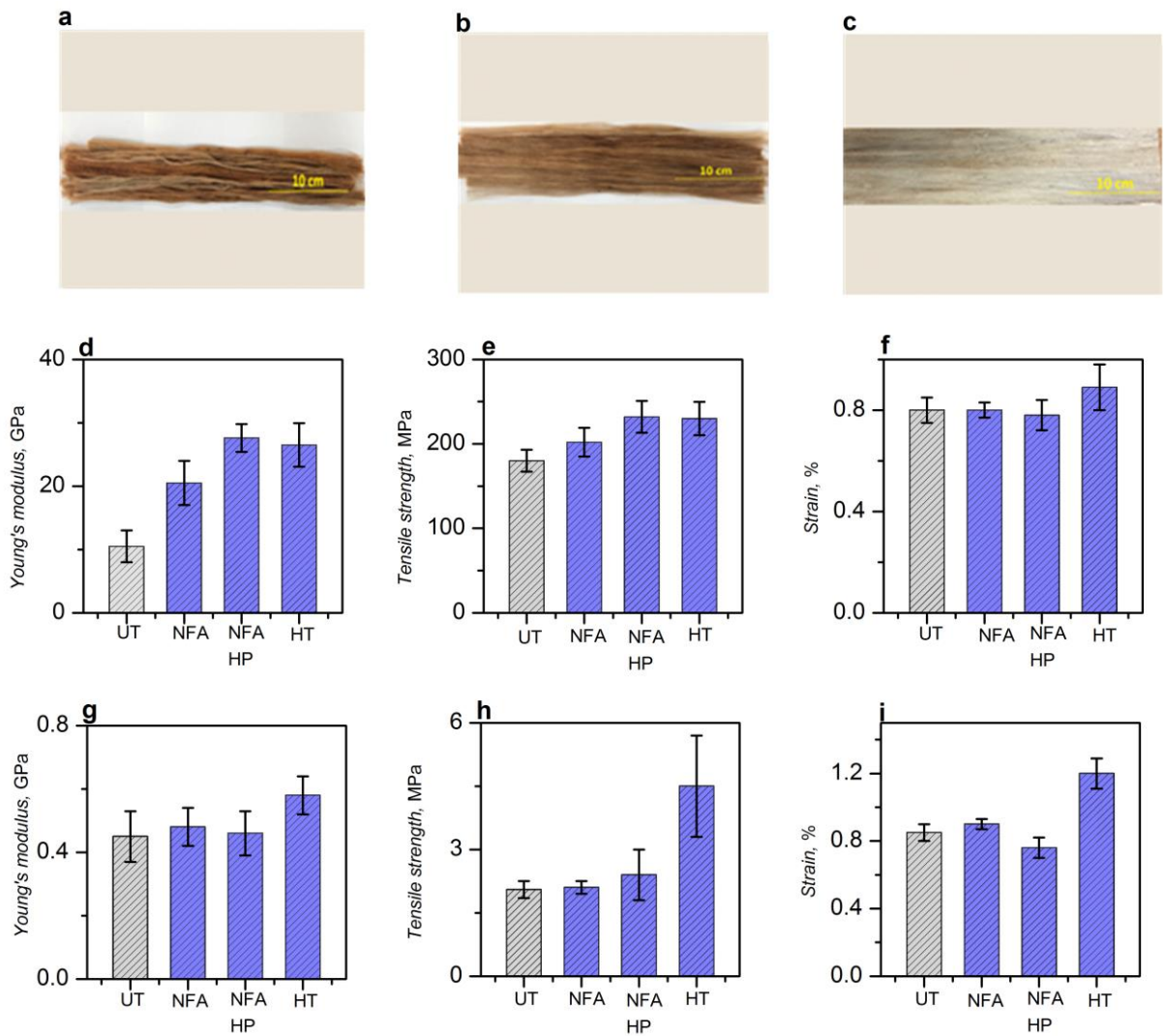

*Figure 2. (a) Untreated fiber, UT preform; (b) fibrillated fiber preform, NFA; (c) fibrillated and compacted preform, NFAHP; (d) longitudinal Young's modulus of untreated and fibrillated jute fiber composites; (e) longitudinal tensile strength of untreated and fibrillated jute fiber composites; (f) longitudinal tensile strain% of untreated and fibrillated jute fiber composites; (g) transverse Young's modulus of untreated and fibrillated jute fiber composites; (h) transverse tensile strength of untreated and fibrillated jute fiber composites; (i) transverse tensile strain% of untreated and fibrillated jute fiber composites.*

We then employ a popular compaction technique with pressing in order to produce NFAWP jute/epoxy composites. This physical treatment increases fiber packing significantly, and also Young's modulus (~27.6 GPa) and tensile strength (~232 MPa) by ~34% and ~15%, respectively, than NFA jute/epoxy composites. The combination of individualisation (combing) and compaction (pressing) improves the packing capacity of the composites significantly and increases their load bearing capacity during tensile tests. Further hot water treatment (HT) increases $V_f$ marginally; however a slight decrease in the stiffness and strength of the composites is observed, may due to the change in the biochemical compositions and macromolecular arrangement.[43]

We also carry out the transverse tensile tests of jute/epoxy composites in order to better understand the effect of NFA and physical treatments on the improvement of interfacial shear strength and mechanical properties. A strong bond is needed for better transverse tensile strengths and even for better water resistance of polymer composites.[44] When compared with longitudinal properties, transverse tensile properties of the composites are found to be lower because of the fiber geometry. Reinforcing fibers are usually distributed parallelly in the direction of loading and hence unable to carry the significant amount of load in the transverse direction as they do in the longitudinal direction. Figure 2 (g-i) show poor transverse properties of the composites with UT jute fibers, due to the presence of impurities (such as fat, waxes, lignin, pectin, and hemi-cellulose) in the fibers that contribute to the poor adhesion between the fiber and matrix. Even with NFA and physical treatments, no improvement in transverse tensile properties is observed. However, after treatment with hot water (HT), transverse tensile strength of HT fiber composites become more than double (~10.6 MPa) than that of UT fibers (~4.16 MPa). This could be explained by the strong interfacial shear strength of HT fibers with epoxy resin than that of UT fibers.[6] This behaviour of the fibers towards epoxy resin gives a preliminary indication to activate the fiber surface either by removing the impurities or filling the flaws to bond or cross-link with resin in order to produce a strong interface.

**Ultra-High Performance of Nano-Engineered Graphene-Based Composites**

Alkali pre-treatment is necessary for jute fibers in order to remove natural impurities and non-cellulosic materials. Moreover, the alkali treatment increases the surface roughness by disrupting hydrogen bonding on the fiber surface. Furthermore, the presence of constituents like hemicellulose can restrict the fiber separation or individualisation, as it is connected with the help of lignin matrix. The alkali treatment removes such constituents (lignin and hemicellulose), and improves fiber packing and fiber-matrix adhesion. The removal of hemicelluloses after alkali treatment was confirmed by using the FTIR *(Figure S9, Supporting Information)*. As shown in Figure 3 (a-c), the Young's modulus increases from ~27.6 GPa to ~32 GPa and the tensile strength increases from ~232 MPa to ~282 MPa after 0.5% alkali treatment of the jute fiber preform (*Table S2, Supporting Information*). Moreover, tensile properties of composites are generally dominated by the fiber properties and fiber orientations.[9] After the mild alkali (HA0.5) treatment, jute fiber surface becomes very clean (*Figure S5b, Supporting Information*), due to the removal of alkali sensitive bonds present between fiber components,[9] which contributes to the better stress transfer between the ultimate cells. Furthermore, the better fiber packing and larger amount of parallel fibers allow composites to bear a higher amount of applied load. Thus, the combination of NFA, Physical and heat-alkali treatments (HA0.5) improves Young's modulus by ~56% and tensile strength by ~56.6% than that of UT jute/epoxy composites.

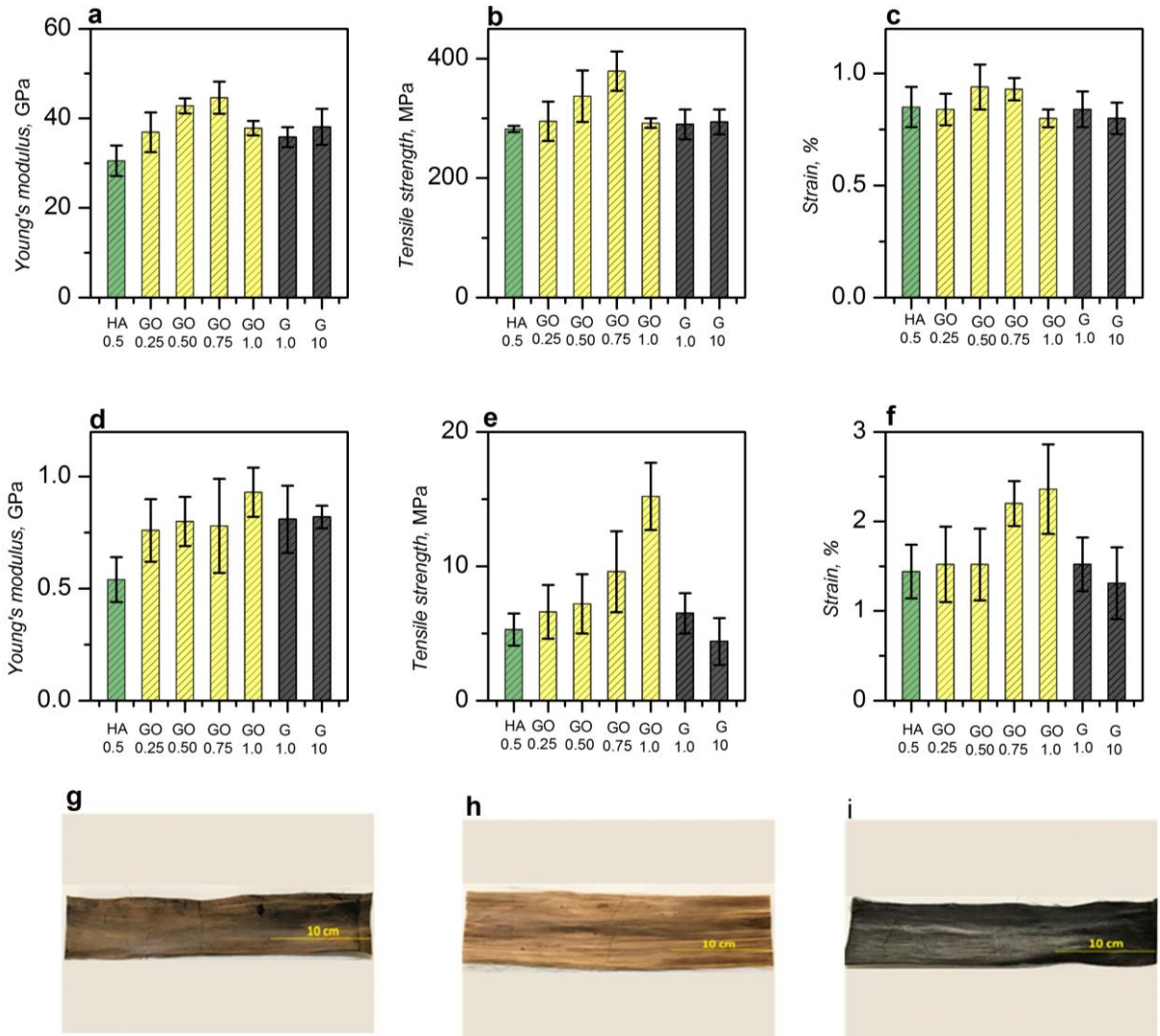

*Figure 3.* *(a) Longitudinal Young's modulus of alkali and graphene jute fiber composites; (b) longitudinal tensile strength of alkali and graphene jute fiber composites; (c) longitudinal tensile strain % of alkali and graphene jute fiber composites; (d) transverse Young's modulus of alkali and graphene jute fiber composites; (e) transverse tensile strength of alkali and graphene jute fiber composites; (f) transverse tensile strain% of alkali and graphene jute fiber composites; (g) alkali treated fiber, HA0.5 preform; (h) GO coated preform, GO0.75; (i) G-flakes coated preform, G1;*

Recent studies have highlighted that the tensile properties of natural fiber can be improved significantly by introducing nano-engineered surface finishes.[7, 14, 18, 45-49] Such studies report that the grafting of nano materials on natural fibers surface increases the surface wettability and roughness, and therefore improves the mechanical properties. Similarly, here we improve the Young's modulus and tensile strength of jute/epoxy composite significantly by nano-engineering of natural fibers surface with graphene materials. The Young modulus, tensile strength and strain% increase with the increase of GO concentration upto 0.75 mg/mL, Figure 3(a-c). At this concentration of GO, we obtain ~39.3% and ~34% increment in young's modulus (~44 GPa) and tensile strength (~379 MPa), respectively, than HA0.5 composites. The combination of physical and chemical treatments of Jute-epoxy composites with NFA and nano surface engineering with GO provides ~324% improvement in the Young's modulus and ~110% tensile strength of the composites.

The enhanced mechanical properties of GO treated jute/epoxy composites could be due to two main reasons: 1) strong adhesion between the GO flakes and HA 0.5 treated fiber and 2) interaction of GO treated jute fiber with the matrix.[6] The oxygen containing functional groups present in GO[41] can create strong bond with the HA treated fibers[50] to make them capable of carrying more load from the matrix.[6] The higher magnification cross-sectional of image of GO treated jute epoxy composites reveals that the elementary fiber which was separated after the alkali and combing process are again strongly connected to each other to produce a strong fiber packing inside the composites. Moreover, we do not observe any porosity related issues in the composites, as GO flakes probably fill those porous spaces. Furthermore, GO contain epoxy groups that could lead to ring-opening reaction to form C-N bonds when exposed to amine groups in epoxy resin.[11] In addition, strong hydrogen bond and mechanical interlocking between GO and epoxy matrix (*Figure S5c and S5c, Supporting Information*) are also possible due to oxygen containing groups and wrinkle structure of GO, respectively.[51]

We obtain 0.75 mg/mL as threshold concentration for GO coating on HA0.5 treated jute fibers, as tensile properties deteriorate after this concentration, may be due to the agglomeration of GO flakes at higher concentration in an epoxy matrix. We then compare the tensile properties of GO treated fiber with G-flakes (G1 and G10) treated jute fiber based jute/epoxy composites. Similar to the single fiber tensile properties,[6] G-flakes does not contribute much to the tensile strength may be due to the absence of functional groups in their structure; but it increases Young's modulus of the composites by ~19% compared to HA0.5 composites, (Figure 3a and *Table S2, Supporting Information*). The improvement in the Young's modulus of the composites might be due to the uniform deposition of large amount of G-flakes on the fiber surface and filling of the micro voids present in HA treated fibers that helps in carrying more amount of load, (*Figure S5d, Supporting Information*).

To better understand the effect of nano-engineering with GO on the interfacial shear strength and mechanical properties of jute/epoxy composites; we carry out the transverse tensile testing of composites, Figure 3(d-f). The treatment with alkali (HA0.5) increases the transverse tensile strength (~5.28 MPa), which is in agreement with previous studies,[52, 53] where they found that alkali treatment improves the transverse tensile strength of natural fiber composites by ~30-150%. Moreover, alkali treatment improves surface wettability and roughness by removing lignin and hemicelluloses; thus enables better penetration of resin and improves the fiber-matrix interaction at the composites interfaces. Figure 3(d-f) shows that the transverse properties of jute/epoxy composites are significantly improved further by GO coating. The transverse tensile strength increases linearly with the increase of GO concentrations, Figure 3e. We achieve maximum transverse tensile strength (~15.26 MPa) with 1 mg/ml GO, which is ~560% and ~189% improvement than the untreated and HA0.5 treated jute fiber epoxy composites, respectively. After the alkali treatment, the hydrogen bonding network in the fiber breaks and the hydroxyl group of cellulose become more active to promote hydrophilicity of the fiber as well as compatibility with the GO sheets. Moreover, GO coated fibers contain a significant amount of oxygen containing functional groups such as hydroxyl (-OH), epoxide (C-O-C), carbonyl (C=O) and carboxyl (O–C=O).[24, 41, 46, 50] Such functional groups interact with the groups of epoxy resin and form a strong mechanical inter-locking at the fiber-matrix interface by a suitable bonding. Moreover, we use an amine-based hardener in order to solidify the fiber/epoxy network, which may form C-N bonds with GO coated fibers through ring opening polymerisation.[18, 45, 51] Therefore, the transverse tensile strength of GO treated jute fiber composites is higher than those of untreated fiber composites. In contrast, the coating with 1 and 10 mg/mL G flakes does not show further noticeable improvement in the transverse tensile strength after the alkali treatment, may be due to the absence of oxygen functional groups in the G flakes. Although lower concentration (~1 mg/mL) of G flakes shows slight improvement up to 6.5 MPa (~ 23 % more) as compared to 10 mg/mL G flakes (only 4% improvement) than HA0.5 treated fibers composite. The lower concentration of G flakes produce better mechanical inter-locking on the fiber surface by the diffusing into alkali treated rough and porous structure of jute fiber (*Figure S5d, Supporting Information*).[6] This results suggests that the GO modified jute fiber could significantly improve the interfacial adhesion between the jute fiber and epoxy matrix.

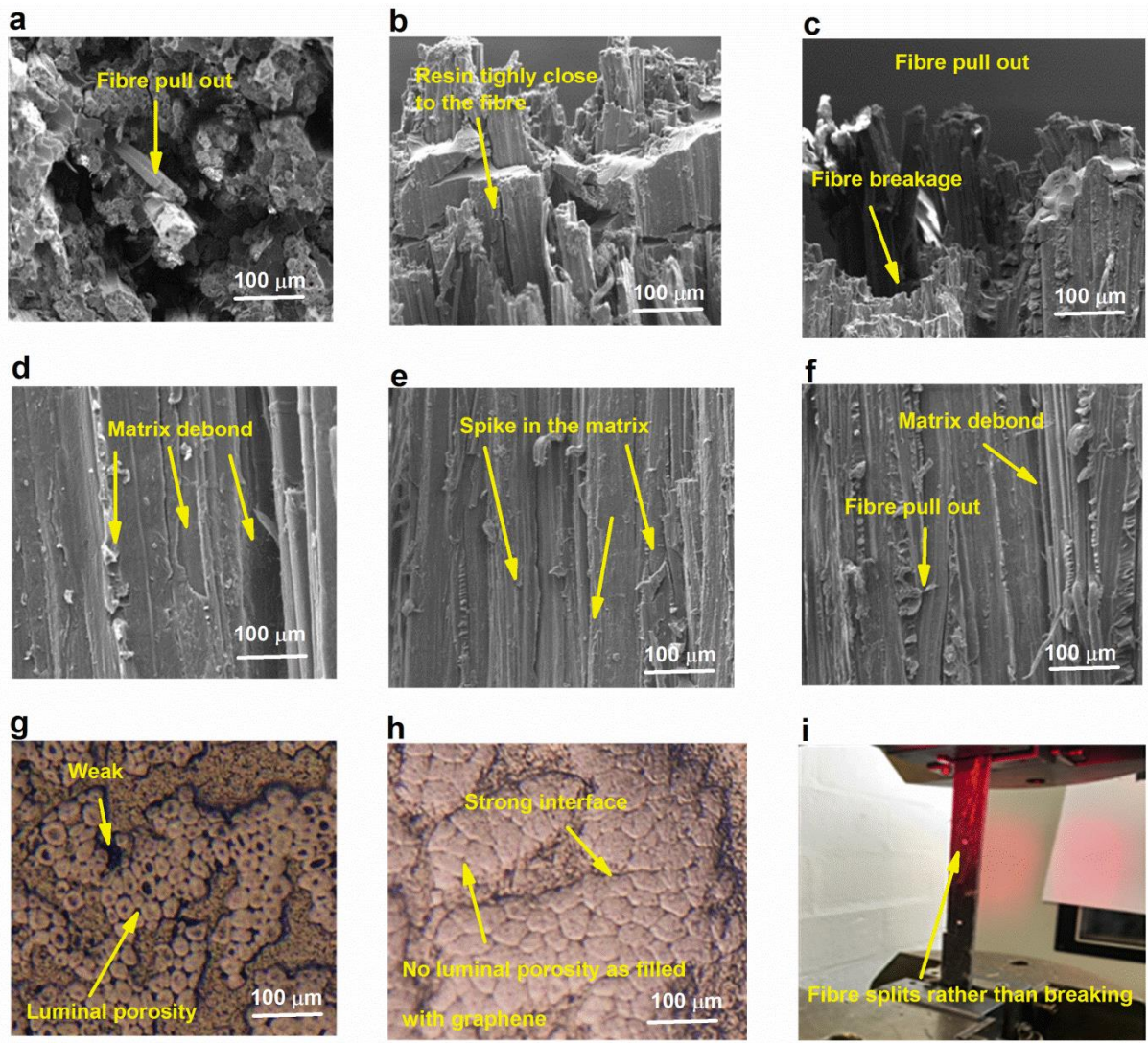

*Figure 4. (a) GO coated specimen after tensile test failed by fiber splitting; (b) fracture surface of UT composites after longitudinal tensile test (X250); (c) fracture surface of GO 0.5 coated composites after longitudinal tensile test (X250); (d) fracture surface of UT composites after transverse tensile test (X250); (e) fracture surface of GO 0.5 coated composites after transverse tensile test (X250); (f) fracture surface of G1 composites after transverse tensile test (X250); (g) Optical microscopic cross-sectional image of alkali treated jute fiber composites with higher magnification (X1000); (h) optical microscopic cross-sectional image of GO coated jute fiber composites with higher magnification (X1000); (i) Fiber splits in the GO coted specimen after the tensile test.*

**Fracture Surface Topography**

We examine the fracture surface of the composites specimen after the longitudinal tensile test using a Scanning Electron Microscope (SEM), Figure 4(a-c). For the untreated jute fiber (Figure 4a), the surface of the composites fails predominantly by the weak interfacial bonding and the fiber pull-out. The uneven fiber breakage occurs along the direction of the fiber. With alkali-treated fibers, the rate of fiber pull-out reduces and more even fiber breakage is observed (*Figure S8a, Supporting Information*). This indicates an improvement in the interfacial bonding between the alkali treated jute fiber and epoxy matrix. Figure 4b shows a modified fracture surface for GO coated jute fibers composites and strong bonding between resin and coated fibers. Moreover, the brittle appearance of matrix (please see yellow arrow mark in the Figure 4b) provides evidence of strong interfacial bonding that can lead to higher tensile properties of the composites. As discussed earlier, GO sheets introduced in the interfacial regions increase the strength and toughness at the interfacial regions due to the 'crack healing' effect[18] and potential chemical bonding between the epoxy and GO sheets. As a result the strong interface of GO modified jute and epoxy matrix transform the failure mode from the fiber pull-out or de-bonding to transverse fracture. Again, G flakes modified jute fiber composites specimen shows more fiber de-bonding (Figure 4c) and pull-out. This may be due to the agglomeration of G flakes at the interfacial regions, which generates various types of stress concentration and thus reduce the strength at the interface.

We also examine the fracture surface of transverse tensile test specimen by SEM in order to better understand the interfacial behaviour and mechanical properties of jute fiber-epoxy composites. The composite fracture section is shown in the Figure 4(d-f). Like longitudinal specimen, the clean and smooth surface of untreated jute fiber reveals poor interfacial bonding of untreated jute fiber-epoxy composites. Figure 4d shows that the matrix is completely detached from the fiber (matrix de-bonding), due to the weak adhesion between the fiber and matrix. It indicates that the fiber matrix adhesion is the dominant mechanism of shear failure with interface being the weakest part of the composites. The fracture surfaces of alkali treated composites shows grooved appearance in the image (*Figure S8b, Supporting Information*), which indicates the improvement of interface after alkali treatment of jute fibers. However, there is still matrix dominating de-bonding area in the fracture surface (*Figure S8b, Supporting Information*). Again for GO modified jute fiber composites, a significant change in the fracture surface are visible (Figure 4e) and we find a high amount of GO flakes sticking to the resin surface. Figure S5c shows the evidence of leaf like flakes on the fiber surface after the transverse tensile test, which creates small spikes on the resin surface (arrow mark in the Figure 4e). This may be the outcomes of strong interaction between the GO modified jute fiber and epoxy network and seems to be responsible for the improvement of interfacial shear strength of the composites. Figure 4f shows a de-boned fracture surface of G flakes modified jute epoxy composites, may be due to the lack of strong bonding between the G flakes and the jute fiber.

We then observe the cross section of alkali and graphene treated composites at higher magnification, to study the porosity of the composites, Figure 4(g, h). Figure 4g represents the interfacial porosity (luminal porosity, impregnation porosity) present in the alkali treated fiber composites. This porosity may also result in the rapid failure of the composites in the transverse direction. Whereas in the Figure 4h, such porosity is not visible as we introduce GO to alkali treated jute fibers. We also examine the failed specimen of all samples after the longitudinal tensile test and find that all the samples fails catastrophically (*Figure S7, Supporting Information*) in the tensile test except GO 0.75 treated jute fiber composites specimen. Figure 4i shows fiber splitting in GO 0.75 composites during the failure of the composites, which provides evidence of strong interface of the composites and in agreement with tensile test results.

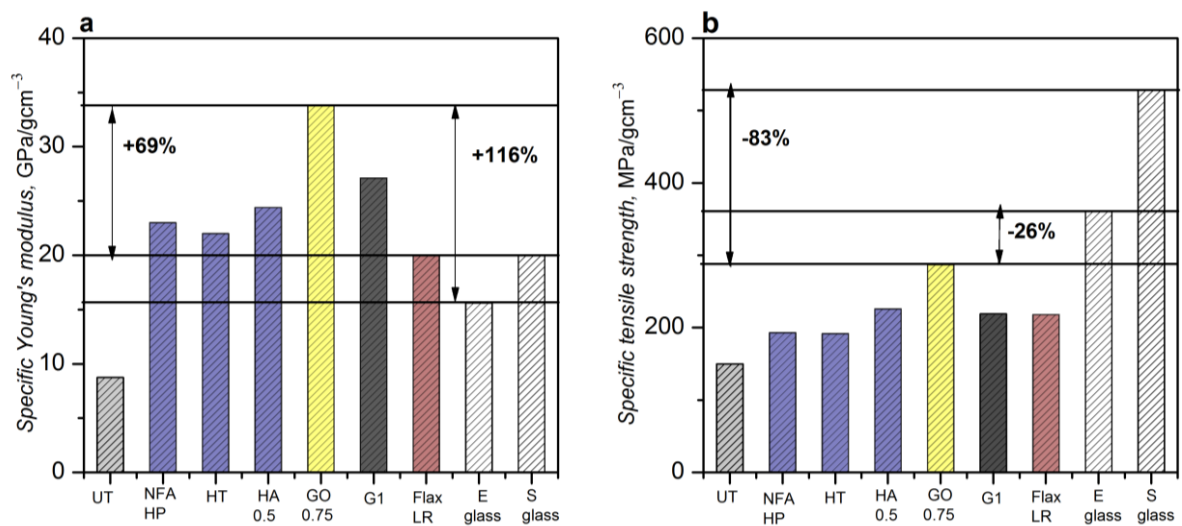

**Figure 5.** Comparative study of specific properties of untreated, new fiber architecture, alkali treated and graphene materials coated composites with Flax, E-glass an S-glass fiber composites: a) Specific Young's Modulus and b) Specific tensile strength. (LR= data taken from the literatures[53])

**Comparative Study**

We also make unidirectional E-glass and S-glass fiber reinforced epoxy composites in order to compare with our obtained results on jute fiber-epoxy composites. We then compare obtained specific properties of untreated, new fiber architecture, alkali treated and graphene materials coated composites with Flax, E-glass an S-glass fiber composites, Figure 5(a-b). The specific Young's modulus of untreated jute fiber composites is found to be ~8.75 GPa/gcm$^{-3}$. Our newly developed jute fiber architecture increases the specific Young's modulus (~23 GPa/ gcm$^{-3}$) of composites significantly, which is higher than that of flax fibers, S-glass and E-glass fiber composites. Please note we use the data from the literature for flax fiber composites with higher volume fraction.[53] After coating jute fibers with GO, the Young's modulus of composites increases significantly to ~33.8 GPa/ gcm$^{-3}$ (Figure 5a), which is ~116% and ~69% higher than that of E (~15.6 GPa/ gcm$^{-3}$) and S (~20 GPa/ gcm$^{-3}$) glass fiber, respectively. Although the specific tensile strength of GO coated composites is found to lower than that of S and E-glass fibers; however the obtained specific tensile strength (~287.29 MPa/ gcm$^{-3}$) with GO coated composites with NFA is found to be higher than any other natural fibers composites.

**Table 1.** Comparing tensile properties of graphene coated jute fiber/epoxy composites with other natural fibers in the literatures along with E and S-glass fiber/epoxy composites.

| Fibers | $V_f$(%) | Young's modulus, GPa | | | Tensile strength, MPa | | | Reference |
|---|---|---|---|---|---|---|---|---|
| | | Before treatment | After treatment | Change % | Before treatment | After treatment | Change % | |
| Jute (NaOH 25%) | 40 | 13.5 | 21.6 | 60 | 100 | 160 | 60 | 9 |
| Flax (NaOH 1%) | 48b-53a | 23 | 25 | 8 | 282 | 283 | 0.35 | 53 |
| Kenaf (NaOH 6%) | 48.6 | 10.34 | 10.7 | 3.5 | 95.4 | 106.3 | 11.5 | 40 |
| Sisal (NaOH 2%) | 50 | 9.5 | 5 | -47.3 | 275 | 320 | 16.3 | 43 |
| Carbon | - | 45 | 55 | 22.2 | 1750 | 2000 | 14.28 | 54 |
| Jute (NaOH 0.5%) | 54 | 27.6 | 32 | 16 | 232 | 282 | 21.5 | **This study** |
| Jute (GO 0.75 %) | 56 | 27.6 | 44.6 | 61.5 | 232 | 379 | 63.3 | **This study** |
| E-glass | 55 | 33.5 | | | 777 | | | **This study** |
| S-glass | 58 | 45 | | | 1187 | | | **This study** |

**\*b - before the test and a- after the test**

We then compare the tensile properties of jute fiber/epoxy composites of this study with other natural fibers based epoxy composites with higher volume fraction reported in the literatures, Table 1. A direct comparison is sometimes difficult as the experimental conditions in those studies are different. In addition, different fibers have different constituents' ratio which have direct impact on the mechanical properties. Table 1 shows a brief comparison between the results obtained in this study with results from previous studies obtained with various surface treatments on jute and other natural fiber based composites. As found in literatures, traditional alkali treatment at lower concentration does not improve mechanical properties significantly. However, the jute fiber treated by GO and graphene flakes in our study shows a fairly a large increment in both Young's modulus (~61.5% for GO) and tensile strength (~63.3% for GO) of the composites than the untreated one.

**Conclusion**

In this work, we report on the individualisation of jute fibers and nano-engineering with graphene oxides and graphene flakes in order to improve the mechanical properties of jute fiber/epoxy composites for high performance natural composites applications. The individualisation of jute fibers by combing improves the fiber packing of the composites significantly and results in a composite with new fiber architecture with higher mechanical performance. Further graphene coating on jute fibers promotes strong interfacial bonding and improved mechanical properties of the composites. Our nano-engineered graphene-based jute fiber composites provide higher mechanical performance and better specific properties than any other natural composites. The Young modulus and tensile strength of jute-epoxy composites is increased by ~324% and ~110%, respectively, more than untreated jute fiber composites. The obtained specific modulus is also higher than that of glass, flax and any other natural fibers composite, and specific tensile strength is comparable to that of E-glass. We believe our graphene–based jute fiber composites with NFA have potential to replace synthetic composites such as glass fibers for stiffness driven structural applications.

**Experimental Methods**

**Materials**

The plant material (*Corchorus Olitorious*) known as "*Tossa white jute*" was obtained from Bangladesh, cultivated on the sandy loam plateau in the Northeast of Dhaka. The sample was cultivated from February to May in 2015. The annual rainfall of this area is ~500-1500 mm and the temperature ranges from 20 to 33 °C. The content of long fibers in the bundles is ~98-99 wt.-%, whereas the rest 1-2 wt.-% is shives (cortical tissues and dust). The untreated long jute fiber has a golden colour with an average length and diameter of ~2.9 m and ~0.059 mm, respectively. Analytical grade sodium hydroxide (NaOH) pellets (product no: 10502731) were purchased from Fisher Scientific, UK. EL2 Epoxy Laminating Resin and AT30 Epoxy Hardener were purchased from Easy Composites, UK. The natural flake graphite (average lateral size 50 mm) was kindly supplied by Graphexel Limited, UK. Sodium deoxycholate (SDC) powder, potassium permanganate (KMnO4), sulfuric acid ($H_2SO_4$, ~99%), and hydrogen peroxide ($H_2O_2$, ~30%) were purchased from Sigma Aldrich, UK. S and E-glass fibers were purchased from AGY, USA. EPI-REZ™ waterborne epoxy resin (product no 7520-W-250) was purchased from Hexion, UK. A modified Hummer's method that was described elsewhere was used to prepare graphene oxide (GO) in water.[55] Our previously reported method was followed to prepare micro-fluidized graphene flakes (G).[28] The lateral dimension of GO and G flakes are ~5.85 μm and ~4.86 μm, respectively; whereas the mean thickness of GO and G flakes are ~2.07 nm and ~2.26 nm, respectively.

**Alkali Treatment**

Untreated jute fibers were washed with deionised (DI) water after cutting into 30 cm long pieces and they were then dried at 80 °C until a constant weight was achieved. These fibers were then treated in warm water at 60 °C for 60 minutes and then boiled at 100 °C for 30 minutes. The weight of fibers was reduced by ~6 wt.-% and labelled as HT fibers. After these cleaning procedures, HT jute fibers were dipped in 0.5 wt.-% NaOH solutions with 1:50 Materials to liquor ratio (M:L) in order to remove hemicelluloses. The fibers obtained after two cycles of alkali treatment is termed as HA0.5 and losses almost similar weight ~6 wt.-% as reported in previous work.[12]

**Manufacturing of Graphene-Based Jute Fiber Preforms with NFA**

A unidirectional jute fiber matt was developed as preform by combining physical and chemical treatments in three steps, (*Figure S1, Supporting Information*). Firstly, a hand combing device was used to separate the elementary fiber from the technical fiber, (*Figure S1a, Supporting Information*). The hand comb was drawn along the length of the fiber for at least twice to get a perfectly aligned and mostly individualised jute fiber, Figure S2b. The both edges of perfectly aligned jute fiber preform was sealed using double-sided tape. Secondly, this preform was then coated with graphene materials (GO and G flakes) for 30 minutes with 1:10 M:L ratio and air dried. The graphene materials coated jute fiber preform was then hand sprayed with 1 wt.-% EPI-REZ™ epoxy solution (binder). Finally, graphene-based performs were hot pressed at 120 °C for 30 minutes at 1 Ton/inch$^2$ pressure and allowed to cool down to get 40x300 cm size preform with NFA, Figure S1f. A very small amount of binder (~0.0015 wt.-% EPI-REZ™) was used in this study, which is epoxy (EL2) compatible with no significant effect on the properties of the resulting composites. We labelled the preforms as follows: untreated preform (UT), new fiber architecture preform after individualisation (combing) (NFA), new fiber architecture preform with pressure (NFAWP), heat alkali treated preform (HA), Graphene oxides treated preform with 0.25, 0.50, 0.75, 1.0 mg/mL concentration were labelled as GO 0.25, GO 0.50, GO 0.75 and GO 1.0 respectively and Graphene flakes treated preform with 1 and 10 mg/mL concentrations were labelled as G 1.0 and G 10. Unidirectional glass fiber preforms were manufactured in order to compare its mechanical properties with that of jute fiber composites from this study (*Figure S4a, Supporting Information*).

**Composite Manufacturing by Vacuum Infusion Process**

A vacuum resin infusion process and room temperature thermoset EL2 epoxy resin was used to manufacture FRC. Briefly, 4 layers of UD jute preforms (dimension: 300 mm x 40 mm) are laid on a pre-cleaned and pre-coated (with PVA release agent) metal plate. The sample was sealed by a plastic bag and vacuumed pressed using a pump. EL2 Epoxy Laminating Resin and AT30 Epoxy Hardener were degassed separately for 30 minutes and mixed together immediately before we use. The resin with hardener was then flown over layered UD jute preforms at a constant flow rate using a vacuum pump, which enabled the impregnation of jute preforms with resin. The resin infused preforms were then cured at room temperature for 24 hours to make jute FRC for further characterization.

**Characterization**

An Optical microscope (Keyence digital microscope VHX-500F, UK) was used to qualitatively measure the image of fiber packing arrangement of the composites (*Figure S2, supporting information*) and flake size of graphene materials. A Philip XL-30 field emission gun scanning electron microscope (SEM) was used to analyse the surface topography of fractured jute fiber composites. The surface characteristics of graphene materials was analysed using a Kratos axis X-ray photoelectron spectroscopy (XPS) system. A Dimension Icon (Bruker) Atomic Force Microscopy (AFM) was used to determine the flake thickness. A Renishaw Raman System equipped with 633 nm laser was used to collect Raman spectra of graphene flakes.

**Volume Fraction and Density**

The fiber volume fraction of laminates was calculated using the ratio of the mass of the preform $W_f$ and the resulting laminate $W_c$. the composite density $\rho_c$ was measured using; the specimen chamber temperature of 20±1 °C was used.

The fiber volume fraction $V_f$, matrix volume fraction $V_m$ and void volume fraction $V_p$ of the composites were calculated using eq 1, where *w and ρ* considered for weight and density, respectively while the subscript *f, m* and *c* denote fibers, matrix and composite, respectively.

$$V_f = \frac{\rho_c}{\rho_f} w_f; V_m = \frac{\rho_c}{\rho_m}(1 - w_f); V_p = \frac{(\rho_{th} - \rho_{exp})}{\rho_{th}} \qquad (1)$$

Density measurements were carried out according to ASTM-D3800-99 in an AJ50L (Mettler Toledo, UK) analytical balance. We weight the untreated and treated composites in the air and then in water. The weight difference between two media is called buoyance force.[56] We calculate the density of the composites by using the following formula eq 2.

$$\rho_c = \frac{M_1}{(M_1 - M_2)} (\rho_l - \rho_{air}) + \rho_{air} \qquad (2)$$

Where, $\rho_l$ is the density of paper oil, $\rho_{air}$ is the density of air, $M_1$ is the weight of sample in air and $M_2$ is the weight of sample in liquid, respectively. For each laminate, a minimum of five samples were tested and the average of five samples was calculated as the final density of the composites.

**Mechanical Testing of the Composites**

The longitudinal tensile test was carried out to determine the tensile properties (Young's modulus and tensile strength) of the unidirectional jute fiber composites. The test was conducted as per ASTM D3039 standard using an Instron 5985 (UK) testing machine equipped with a 100 kN load cell and a video extensometer (Figure S6). Five 250-mm long and 15-mm wide specimen were tested for each type of composites at a cross-head speed of 2 mmm/min. The Young's modulus, tensile strength and tensile failure strain were measured from obtained stress-strain curve. In addition, transverse tensile test was carried out to understand the effect of GO and G flakes on the mechanical properties of the composites.

**Associated Contents**

**Supplementary Information**

Supporting information contains the design, manufacturing of jute fiber preform and its composites; preparation of jute and glass fiber preform; single fiber properties; microscopy and density of the composites; tensile test and mechanical properties of the composites; optical and SEM images of fracture surface of jute-epoxy composites with untreated, treated and coated samples.

**Contributions**

N.K. and F.S. conceived and designed the project. F.S. prepared the samples, performed the measurements, and carried out data analysis under supervision of P.P. F. S. and N.K. wrote the manuscript with K.S.N. S.A. prepared, characterized and analysed graphene materials. All other authors contributed to the discussion of the manuscript.

**Corresponding Authors**

All the correspondence should be addressed to mdnazmul.karim@manchester.ac.uk and prasad.potluri@manchester.ac.uk

Acknowledgments

Authors kindly acknowledge Commonwealth Scholarship Council, U.K. and the Government of Bangladesh for the PhD funding of Forkan Sarker and Shaila Afroj, respectively. This work was supported by EU Graphene Flagship Program, European Research Council Synergy Grant Hetero2D, the Royal Society, and Engineering and Physical Sciences Research Council, U.K. (EPSRC Grant Number: EP/N010345/1, 2015).

Notes

The authors declare no competing financial interest.